\documentclass[5p,final]{elsarticle}
\usepackage{graphicx,bm,amssymb,amsmath}
\journal{Physics Letters B}
\begin{document}
\begin{frontmatter}

\title{Effect of collective neutrino flavor oscillations on  $\nu p$-process
nucleosynthesis}

\author[gsi]{G.~Mart{\'i}nez-Pinedo}
\author[gsi]{B.~Ziebarth}
\author[gsi,tu]{T.~Fischer}
\author[gsi,tu,fias]{K.~Langanke}

\address[gsi]{GSI Helmholtzzentrum f\"ur Schwerionenforschung,
   Planckstra{\ss}e~1, 64291 Darmstadt, Germany}
\address[tu]{Technische Universit{\"a}t Darmstadt,
  Schlossgartenstra{\ss}e 9, 64289 Darmstadt, Germany} 
\address[fias]{Frankfurt Institute for Advanced Studies, Ruth-Moufang
  Stra{\ss}e 1, Frankfurt, Germany} 

\date{\today}

\begin{abstract}
  The $\nu p$ process is a primary nucleosynthesis process which
  occurs in core collapse supernovae. An essential role in this
  process is being played by electron antineutrinos. They generate, by
  absorption on protons, a supply of neutrons which, by $(n,p)$
  reactions, allow to overcome waiting point nuclei with rather long
  beta-decay and proton-capture lifetimes.  The synthesis of heavy
  elements by the $\nu p$ process depends sensitively on the
  $\bar{\nu}_e$ luminosity and spectrum. As has been shown recently,
  the latter are affected by collective neutrino flavor oscillations
  which can swap the $\bar {\nu}_e$ and $\bar{\nu}_{\mu,\tau}$ spectra
  above a certain split energy.  Assuming such a swap scenario, we
  have studied the impact of collective neutrino flavor oscillations
  on the $\nu p$-process nucleosynthesis. Our results show that the
  production of light $p$-nuclei up to mass number $A=108$ is very
  sensitive to collective neutrino oscillations.
\end{abstract}

\begin{keyword}
explosive nucleosynthesis \sep $\nu p$ process \sep core collapse supernova \sep
collective neutrino oscillations 
\PACS 26.30.$-$k, 97.60.Bw 
\end{keyword}
\end{frontmatter}

Core-collapse supernova explosions have been identified as the site
for explosive nucleosynthesis~\cite{Janka.Langanke.ea:2007}.  Of
particular interest here is matter that is ejected from the surface of
the newly born neutron star due to late-time neutrino heating, known
as the neutrino-driven wind. It develops on timescales of seconds
after the onset of the explosion.  Due to the high temperatures at the
surface, this matter is ejected as free protons and neutrons.  Upon
reaching cooler regions with increasing distance from the neutron
star, nucleons can be combined into nuclei.  The outcome of this
nucleosynthesis depends on the proton-to-baryon ratio of the ejected
matter.  This ratio is determined by the competition of electron
neutrino and antineutrino absorption on nucleons and their inverse
reactions,

\begin{equation}
\nu_e + n \rightleftarrows p + e^-, \qquad
\bar{\nu}_e + p \rightleftarrows n + e^+.
\label{eq:1}
\end{equation}
It strongly depends on the neutrino and antineutrino luminosities and
spectra.  Simulations~\cite{Liebendoerfer.Mezzacappa.ea:2003,%
  Buras.Rampp.ea:2006,Huedepohl.ea:2010,Fischer.Whitehouse.ea:2010}
indicate that in an early phase of the explosion, the ejected matter
is proton-rich; i.e. the proton-to-neutron ratio $Y_e$ is larger than
0.5.  When this proton-rich matter expands and cools, nuclei can form
resulting in a composition dominated by $N=Z$ nuclei, mainly $^{56}$Ni
and $^4$He, and protons~\cite{Seitenzahl.Timmes.ea:2008}.  Without the
further inclusion of neutrino and antineutrino reactions the
composition of this matter will finally consist of protons,
alpha-particles, and heavy (Fe-group) nuclei (in nucleosynthesis terms
a proton- and alpha-rich freeze-out), with enhanced abundances of
$^{45}$Sc, $^{49}$Ti, and
$^{64}$Zn~\cite{Froehlich.Hauser.ea:2006,Pruet.Woosley.ea:2005},
solving the longstanding puzzle of underproduction of these elements
in stellar nucleosynthesis studies.  The matter flow stops at $A=64$,
due to the long beta-decay half-live of $^{64}$Ge, 63.7(25)~s, which is much
longer than the expansion time scale, and the small proton separation
energy of $^{65}$As, $-90(85)$~keV~\cite{Tu.Xu.ea:2011}, which
suppresses proton capture on $^{64}$Ge at the high temperatures
involved.

However, neutrino reactions play an essential role during the
nucleosynthesis.  $N \sim Z$ nuclei are practically inert to neutrino
captures (which convert a neutron into a proton), because such
reactions are endoergic for neutron-deficient nuclei located away from
the valley of stability.  The situation is different for electron
antineutrinos that are captured in a typical time of a few seconds,
both on protons and nuclei, at distances of $\sim 500$~km.  This time
scale is much shorter than the beta-decay half-lives of the most
abundant heavy nuclei reached without neutrino interactions (e.g.\
$^{56}$Ni, $^{64}$Ge).  As protons are more abundant than heavy
nuclei, electron antineutrino capture occurs predominantly on protons,
causing a residual density of free neutrons of
$10^{14}$--$10^{15}$~cm$^{-3}$ for several seconds, when the
temperatures are in the range of 1--3~$\times 10^9$~K.  These neutrons
allow now the matter flow to overcome waiting point nuclei like
$^{64}$Ge via $(n,p)$ reactions and to continue to heavier nuclei.
The associated nucleosynthesis process is called the $\nu
p$-process~\cite{Froehlich.Martinez-Pinedo.ea:2006,%
  Pruet.Hoffman.ea:2006,Wanajo:2006}.

\begin{figure}
  \centering
  \includegraphics[width=\linewidth]{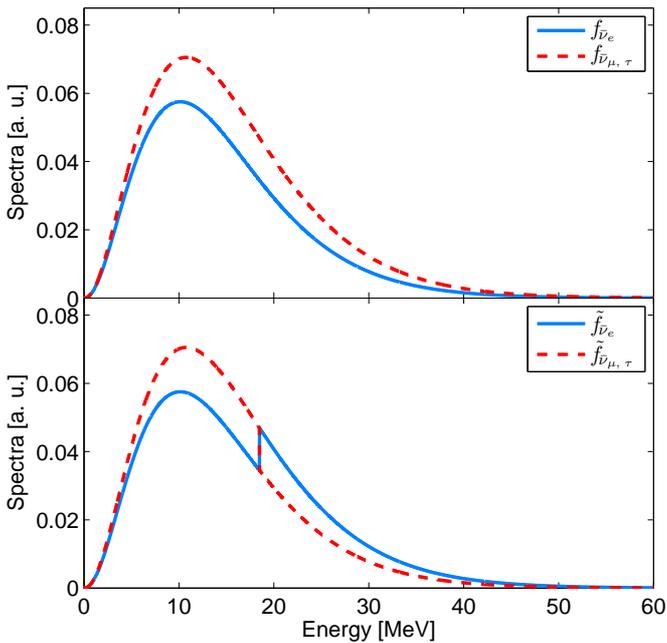}
  \caption{upper panel: $\bar{\nu}_e$ and $\bar{\nu}_{\mu,\tau}$ spectra
    $f_{\bar\nu_e}$ and $f_{\bar\nu_{\mu,\tau}}$ based on the
    simulations of ref.~\cite{Buras.Rampp.ea:2006} as given in table 1
    of ref.~\cite{Pruet.Hoffman.ea:2006}. The ${\bar\nu_e}$ spectra
    have been arbitrarily normalized to one while the ${\bar
      \nu_{\mu,\tau}}$ spectra are normalized to 1.3 to keep the
    relative ratio of the number luminosities. lower part: modified
    spectra $\tilde {f}_{\bar \nu_e}$ and $\tilde {f}_{\bar
      \nu_{\mu,\tau}}$, including the effects of collective neutrino
    oscillations, which induce a swap of the ${\bar \nu_e}$ and ${\bar
      \nu_{\mu,\tau}}$ spectra for energies above $E_s \approx 18$~MeV
    based on the normal mass herarchy calculations of
    ref.~\cite{Dasgupta.Dighe.ea:2009} .\label{fig:spectra}}
\end{figure}

How far the massflow can proceed in the $\nu p$ process strongly
depends on the environment conditions. In particular, the sensitivity
to the temperature at neutrino-driven wind termination and the $Y_e$
value of the ejected matter has been investigated in
ref.~\cite{Wanajo.Janka.Kubono:2011}. Here we explore the sensitivity
to the luminosity and spectrum of electron antineutrinos. These are
expected to be affected by collective neutrino flavor oscillations
that occur in the high-neutrino-density environment surrounding the
neutron star, see~\cite{Duan.Fuller.Qian:2010} for a recent
review. Several studies~\cite{Dasgupta.Dighe.ea:2009,%
Duan.Friedland:2011,Galais.Volpe:2011,Wu.Qian:2011} have shown that
collective neutrino oscillations swap the spectra of $\bar{\nu}_e$ and
$\bar{\nu}_{\mu,\tau}$ neutrinos in certain energy intervals bounded
by sharp spectral splits. The split energy depends on the relative
fluxes of $\bar{\nu}_e$ and $\bar{\nu}_{\mu,\tau}$ and on the neutrino
mass hierarchy~\cite{Fogli.Lisi.ea:2009,Mirizzi.Tomas:2010}.  In the
following, as illustrated in Fig.~\ref{fig:spectra} we assume that
collective neutrino flavor oscillations exchange the spectra of
$\bar{\nu}_{e}$ and $\bar{\nu}_{\mu,\tau}$ above a certain split
energy as found by Dasgupta \emph{et
  al.}~\cite{Dasgupta.Dighe.ea:2009} in the case of normal mass
hierarchy.

In general, $\bar{\nu}_{\mu,\tau}$ have a larger mean energy and a
larger high-energy tail than $\bar{\nu}_e$.  This is caused by the
fact that, besides their interactions by neutral-current reactions,
$\bar{\nu}_e$ also interact with the dense neutron star matter by
charge current reactions (which is energetically not possible for
supernova $\bar{\nu}_{\mu,\tau}$ antineutrinos) and hence decouple at
slightly larger radii and lower temperatures.  Due to the high-energy
enhancement of ${\bar{\nu}_e}$, collective neutrino flavor
oscillations are expected to increase the neutron production rate by
antineutrino absorption on protons during the $\nu p$ process and
hence its efficiency to synthesize heavy elements.

We assume that collective neutrino flavor oscillations do not change
the $Y_e$ value which is determined by electron neutrino and
antineutrino absorptions on nucleons close to the neutron star
surface. Furthermore, we assume that collective neutrino oscillations
occur before the onset of $\nu p$-process nucleosynthesis at distances
of $\sim 500$~km. These assumptions are consistent with the recent
multi-angle calculations of
ref.~\cite{Duan.Friedland:2011}. Furthemore, there are no free
neutrons present and $N \sim Z$ nuclei are practically innert to
neutrino absorptions. Hence, the impact of collective neutrino flavor
oscillations on $\nu p$-process nucleosynthesis is to modify the rate
of antineutrino absorption on protons and hence the neutron production
rate.

The relevant absorption rate at distance $r$ from the neutron star
center is defined as follows,
\begin{equation}
\lambda_{\bar{\nu}_e}= \frac{1}{4 \pi r^2} \int_0^\infty dE
\sigma_{\bar\nu_e} (E) f_{\bar\nu_e}(E)  
\label{eq:rate},
\end{equation}
with $\sigma_{\bar\nu_e} (E)$ the absorption cross section for
antineutrinos of energy $E$. For the neutrino spectra we adopt the
following distribution~\cite{keil.raffelt.janka:2003}:

\begin{align}
  f(\alpha,E) = \frac{L_n}{\Gamma(1+\alpha)} \left(\frac
    {1+\alpha}{\langle E \rangle}\right)^{1+\alpha} E^\alpha
  \exp\left(-\frac{(1+\alpha) E}{\langle E \rangle}\right), \label{eq:alphas}
\end{align}
with the neutrino number luminosity, 
\begin{equation}
  \label{eq:ln}
  L_n = \int_0^\infty dE f(\alpha, E)
\end{equation}
and $\langle E \rangle$ the mean neutrino energy.
The parameter $\alpha$ is fixed by the relation:

\begin{align*}
\langle E^2 \rangle = \frac{\alpha+2}{\alpha+1} \langle E \rangle^2.
\end{align*}
It has been shown that such an $\alpha$ distribution gives a good description
of the neutrino spectra calculated in supernova
simulations~\cite{keil.raffelt.janka:2003}. 

With the assumptions above, the $\bar{\nu}_e$ spectrum,
$\tilde{f}_{\bar\nu_e}$, modified by collective neutrino 
oscillations is

\begin{align}
\tilde {f}_{\bar \nu_e} (E) =
\begin{cases}
 f_{\bar{\nu}_e}(E),& E<E_s \\
 f_{\bar\nu_{\mu,\tau}}(E), & E>E_s. 
\end{cases}
\end{align}
According to ref.~\cite{Dasgupta.Dighe.ea:2009}, we expect the split energy,
$E_s$, to be around 18~MeV (see Fig.~\ref{fig:spectra}).

\begin{figure}
  \centering
  \includegraphics[width=\linewidth]{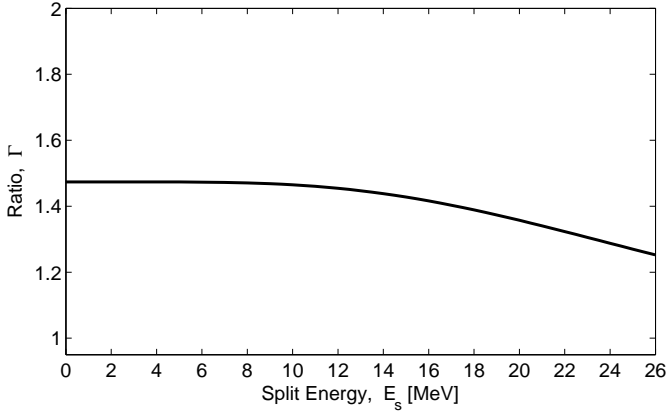}
  \caption{Relative change $\Gamma$ of the neutron production rate
    with respect to the split energy $E_s$, calculated for the
    antineutrino spectra shown in the upper part of
    Fig.~\ref{fig:spectra}. \label{fig:Gamma}}
\end{figure}

The modification of the antineutrino spectrum will affect the neutron
production rate by antineutrino absorption on protons, which in turn
will alter the $\nu p$-process nucleosynthesis. To quantify these
effects, we define the change in the neutron production rate as:

\begin{equation}
\Gamma=\frac{
\int_0^\infty \sigma_{\bar \nu_e} (E) \tilde {f}_{\bar \nu_e}(E) dE}
{\int_0^\infty \sigma_{\bar \nu_e} (E) f_{\bar \nu_e}(E) dE}.
\label{eq:Gamma}  
\end{equation}
To calculate the $\Gamma$ factor we approximate the antineutrino
absorption cross section by 

\begin{equation}
\label{eq:cross}
\sigma_{\bar\nu_e}(E) =
\begin{cases}
  0, & E< \Delta \\
 9.3 \times 10^{-44} \left(\frac{E - \Delta}{\text{MeV}}\right)^2
 \text{cm}^2, & E > \Delta   
\end{cases}
\end{equation}
where $\Delta =1.293$~MeV is the neutron-proton mass difference.
This approximation suffices to calculate $\Gamma$. However, for the
nucleosynthesis studies we use a cross section which also considers
weak magnetism and nucleon recoil corrections~\cite{Horowitz:2002}

Our nucleosynthesis calculations are based on the supernova
simulations of a 15~M$_\odot$ star~\cite{Buras.Rampp.ea:2006} and its
associated nucleosynthesis~\cite{Pruet.Woosley.ea:2005,%
Pruet.Hoffman.ea:2006}.  In particular, luminosities and spectra
parameters for all neutrino flavors are given in table 1 of
ref.~\cite{Pruet.Hoffman.ea:2006}.  We have approximated these spectra
by an $\alpha$-distribution, see Eq.~(\ref{eq:alphas}).  For the
${\bar \nu_e}$ spectrum we find the parameter $\alpha_{\bar \nu_e}
=2.3$ and an average neutrino energy $\langle E_{\bar \nu_e} \rangle =
14.56$~MeV, while for the ${\bar \nu_{\mu,\tau}}$ flavor these
parameters are $\alpha_{\bar \nu_{\mu,\tau}} =2.3$ and $\langle
E_{\bar\nu_{\mu,\tau}} \rangle = 15.44$~MeV.  These spectra $f_{{\bar \nu_e}}
(E)$ and $f_{{\bar \nu_{\mu,\tau}}} (E)$ are plotted in the upper part
of Fig.~\ref{fig:spectra}. The lower panel shows the modified
$\bar{\nu}_e$ spectrum including the effect of collective neutrino
oscillations.  We observe the increased flux of $\bar{\nu}_e$
neutrinos with $E>E_s$ in the modified spectrum.  Due to the energy
dependence of the neutrino absorption cross section, see
Eq.~(\ref{eq:cross}), this increase of high-energy neutrinos will
enhance the neutron production rate.  This is demonstrated in
Fig.~\ref{fig:Gamma} that shows the neutron production rate to be
increased by a factor $\Gamma \approx 1.4$ due to collective neutrino
oscillations.  Importantly, we also observe that $\Gamma$ is
relatively insensitive to the unknown split energy $E_s$, in the range
up to 25~MeV.  This allows us to describe the effect of collective
neutrino oscillations in $\nu p$-process nucleosynthesis studies by
scaling the antineutrino absorption rate by a constant factor
$\Gamma$. We have considered this scaling for radii larger than 500
km, corresponding to temperatures smaller than 3~GK at which the $\nu
p$ process operates.

\begin{figure}
  \centering
  \includegraphics[width=\linewidth]{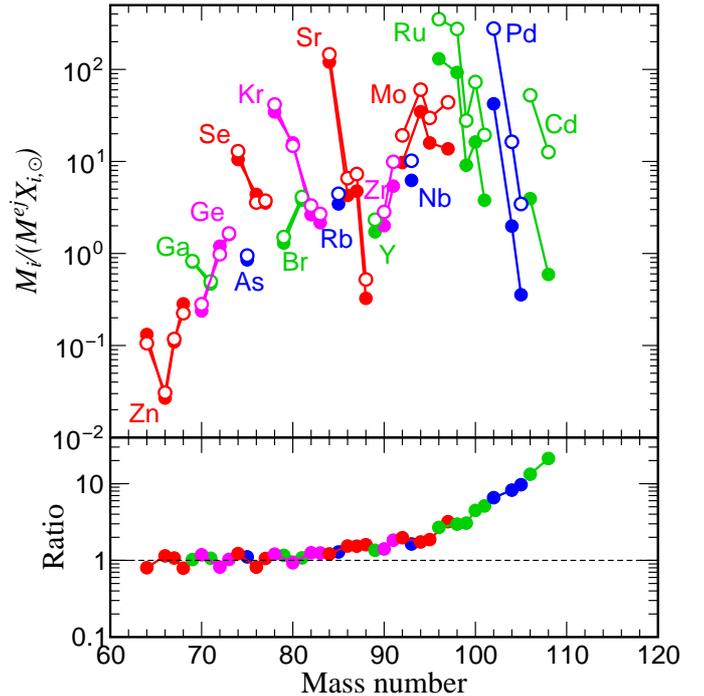}
  \caption{Comparison of overproduction factors calculated in $\nu
    p$-process nucleosynthesis studies with $\Gamma=1.4$, hence
    considering collective neutrino oscillations, and
    $\Gamma=1.0$. The lower panel shows the ratio of overproduction
    factors for the two nucleosynthesis studies as function of mass
    number.  \label{fig:abund} }
\end{figure}

Adopting $\Gamma=1.4$ from Fig.~\ref{fig:Gamma} we have performed a
nucleosynthesis calculation using a sufficiently large nuclear network
considering rates for reactions mediated by the strong,
electromagnetic and weak interaction and involving nuclei in the mass
range up to the europium isotopes (see
ref.~\cite{Arcones.Martinez-Pinedo:2011} for additional details).  The
evolution of temperature, density and $Y_e$ corresponds to the
trajectory labelled ``1116 ms'' shown in figure~3 of
ref.~\cite{Pruet.Woosley.ea:2005}.  The abundance distributions of
elements heavier than $A=64$ are compared to those obtained in a
calculation, in which we choose $\Gamma=1.0$ and kept all other
quantities the same.  This study hence corresponds to a standard $\nu
p$-process calculation without consideration of collective neutrino
oscillations.  In Fig~\ref{fig:abund}, we show the calculated
`overproduction factors' for both nucleosynthesis studies defined by
the ratio $M_i/(M^{\text{ej}} X_{i,\odot})$, where $M_i$ is the
produced mass of isotope $i$ and $X_{i\odot}$ is its solar mass
fraction.  The total mass ejected in the supernova simulation, $M^{\rm
  ej}$, is taken from~\cite{Pruet.Woosley.ea:2005}.  The enhanced
neutron production due to collective neutrino oscillations has two
interesting consequences.  Firstly, it increases the abundances of
nuclei heavier than $A=80$.  In particular, the abundances of light
$p$-nuclides ($^{92,94}$Mo, $^{96,98}$Ru), whose production might be
attributed to the $\nu p$ process, are enhanced by factors 2--3.  The
increase becomes more significant for nuclides with $A>96$. This is
due to the fact that the enhanced rate for neutron production
increases the number of neutrons that can induce $(n,p)$ reactions on
heavy nuclei.  In particular, it reduces the timescale of this
reaction on the $N=50$ nucleus $^{96}$Pd which acts like a ``seed''
for the production of nuclei with
$A>96$~\cite{Wanajo.Janka.Kubono:2011}. Our results indicate that, in
addition to the uncertainties in the supernova dynamics and the
nuclear physics input discussed in
ref.~\cite{Wanajo.Janka.Kubono:2011}, collective neutrino oscillations
play a very important role to determine if the $\nu p$ process can
contribute to the solar inventory of light $p$-nuclei up to $A=108$
($^{108}$Cd). The production of heavier $p$-nuclei must be likely
attributed to other nucleosynthesis processes.

The overproduction factors for nuclei with mass numbers $A=64,68,72$,
and 76 are slightly reduced in a $\nu p$ process which considers
collective neutrino oscillations.  This is due to the fact that their
progenitors, the $\alpha$-nuclei like $^{64}$Ge or $^{68}$Se, serve as
waiting points in the mass flow to heavier elements.  As these waiting
points are overcome by $(n,p)$ reactions within $\nu p$-process
nucleosynthesis, the availability of a larger amount of free neutrons
during the nucleosynthesis process makes the lifetimes of the $\alpha$
nuclei against $(n,p)$ reactions shorter.  As a consequence, less
matter is halted at the waiting points, reflected in the smaller
abundances of nuclides like $^{64}$Zn and $^{68}$Zn, which are
produced via the $\alpha$-nuclei progenitors.

Recent supernova
simulations~\cite{Huedepohl.ea:2010,Fischer.Whitehouse.ea:2010} show
that the $\bar \nu_e$ and $\bar \nu_{\mu,\tau}$ spectra become
increasingly similar with time after the onset of the explosion, being
identical after around
10~s~\cite{Fischer.Martinez-Pinedo.ea:2011}. Additionally, it has been
recently shown that collective neutrino flavor oscillations are
suppressed during the early accretion phase of the supernova
explosion~\cite{Chakraborty.Fischer.ea:2011}. Consequently, we expect
a reduced impact of collective neutrino oscillations on
nucleosynthesis studies both at early and late times. We also find
smaller $\Gamma$ factors if we calculate the enhancement in the
neutron production ratio using the neutrino spectra published for the
simulations of a ONeMg core 8.8~M$_\odot$
star~\cite{Huedepohl.ea:2010,Fischer.Whitehouse.ea:2010} and a
iron-core 10.8~M$_\odot$ star~\cite{Fischer.Whitehouse.ea:2010}.  In
both cases, $\Gamma$ is larger than 1, i.e. collective neutrino
oscillations lead to increased neutron production but the obtained
values, around 1~s after the explosion, $\Gamma=1.1$ (8.8~M$_\odot$
star) and 1.3 (10.8~M$_\odot$ star) are smaller than discussed above.
We have repeated the nucleosynthesis calculations with these $\Gamma$
factors.  We observe that the overproduction factor of the elements
with $A>80$ grows nearly linear with $\Gamma$.  For example, the
overproduction factor of $^{96}$Ru is 2.3 for $\Gamma=1.3$, while it
is nearly 3 for $\Gamma=1.4$.

We have studied $\nu p$-process nucleosynthesis considering the impact
of collective neutrino flavor oscillations. In accordance with recent
studies we assume that these result in a complete swap in the spectra
of electron and non-electron flavors above a certain split
energy. This swap results in enhanced electron antineutrino
captures. Adopting antineutrino spectra from recent supernova
simulations we find that this enhanced capture increases the amount of
free neutrons, present during $\nu p$-process nucleosynthesis, by
20--40\%. This larger supply of neutrons boosts the matter flow to
heavier nuclides and results in larger abundances of nuclides with
$A>64$.  Our current study makes the simplifying assumption that the
collective neutrino flavor oscillations occur before the onset of $\nu
p$ nucleosynthesis which allows to approximate their effect by a
constant increase of the neutron production rate. Due to the
intriguing effect of the oscillations on the production of light $p$
nuclides, we are planning to improve on our current study.  We will
perform nucleosynthesis calculations that consistently include
collective neutrino flavor oscillations based on neutrino spectra and
trajectories from simulations of supernova explosions.

\section*{Acknowledgements}
\label{sec:acknowledgements}

The work has been initiated during the GSI Summer School Program.
B. Ziebarth thanks the Program for financial support. We thank
Y.-Z. Qian and C. Volpe for useful discussions on collective neutrino
oscillations. We also like to acknowledge support from HIC for FAIR,
the Helmholtz Alliance EMMI and the SFB 634 at the Technical
University Darmstadt.


\end{document}